\documentclass[twocolumn,english,pre]{revtex4-1}
\usepackage[T1]{fontenc}
\usepackage[latin9]{inputenc}
\usepackage{color}
\usepackage{amsmath}
\usepackage{amssymb}
\usepackage{graphicx}
\usepackage{esint}

\makeatletter

\@ifundefined{textcolor}{}
{%
 \definecolor{BLACK}{gray}{0}
 \definecolor{WHITE}{gray}{1}
 \definecolor{RED}{rgb}{1,0,0}
 \definecolor{GREEN}{rgb}{0,1,0}
 \definecolor{BLUE}{rgb}{0,0,1}
 \definecolor{CYAN}{cmyk}{1,0,0,0}
 \definecolor{MAGENTA}{cmyk}{0,1,0,0}
 \definecolor{YELLOW}{cmyk}{0,0,1,0}
}

\makeatother

\usepackage{babel}
\begin{document}

\title{Noise, diffusion, and hyperuniformity }

\author{Daniel Hexner\textsuperscript{1,2} and Dov Levine\textsuperscript{1}}

\affiliation{\textsuperscript{}1. Department of Physics, Technion, Haifa 32000,
Israel}

\affiliation{2. James Franck Institute, The University of Chicago, Chicago, Illinois
60637, USA}
\begin{abstract}
We consider driven many-particle models which have a phase transition
between an active and an absorbing phase. Like previously studied
models, we have particle conservation, but here we introduce an additional
symmetry - when two particles interact, we give them stochastic kicks
which conserve center of mass. We find that the density fluctuations\textcolor{black}{{}
in the active phase}\textcolor{red}{{} }\textcolor{black}{decay} in
the fastest manner possible for a disordered isotropic system, and
we present arguments that the large scale fluctuations are determined
by a competition between a noise term which generates fluctuations,
and a deterministic term which reduces them. Our results may be relevant
to shear experiments and may further the understanding of hyperuniformity
which occurs at the critical point.
\end{abstract}


\maketitle
Among the remarkable behavior exhibited by driven many-body systems
is a non-equilibrium phase transition between phases with different
dynamics. For one interesting class of systems, the transition is
from an active phase, where the system evolves in time forever, to
an absorbing phase, where the dynamics eventually ceases. Such behavior
has been studied theoretically\cite{hinrichsen_non-equilibrium_2000,Lubek},
as well as in experiments on sheared particles\cite{pine_chaos_2005,corte,Bartolo1,Bartolo2}.
In all cases, the phase transition is effected by changing a control
parameter such as the density. 

The theoretical models showing this behavior consist of particles,
each of which may be either static or active, depending on its local
neighborhood. For example, in the random organization (RandOrg) model,
initially introduced in \cite{corte} to model experiments on periodically
sheared Brownian particles\cite{pine_chaos_2005}, particles of unit
radius are placed in a volume. If two particles overlap, they are
considered active. At every time step, each active particle is given
a random displacement, or ``kick'', while the isolated inactive
particles remain in place. After an initial transient, and depending
on the value of the control parameter $\phi$ (here, the density)
of the model, the system evolves either into an absorbing state consisting
only of isolated particles, or into an active phase where a well-defined
fraction of the particles are active, and undergo unceasing random-like
motion. This model has been shown\cite{Ramaswamy} to belong to a
larger class of absorbing phase transitions called the Manna universality
class\cite{Non-Equilibrium_Book}.

Although the system is out of equilibrium, the transition between
active and absorbing phases has properties of a continuous phase transition,
with characteristic critical exponents\cite{Non-Equilibrium_Book}
and a well-defined critical value, $\phi_{c}$, of the control parameter.
Thus, the fraction of active particles grows as a power-law $\rho_{a}\propto\left|\phi-\phi_{c}\right|^{\beta}$
, the time scale to reach an absorbing state (or the active steady
state for $\phi>\phi_{c}$) diverges as $\tau\propto\left|\phi-\phi_{c}\right|^{-\nu_{\parallel}}$,
and an appropriately defined correlation length%
\footnote{$\xi$ may be defined in different ways, such as the spreading length\cite{Non-Equilibrium_Book}
or a crossover length\cite{PRL_hyper}.%
} diverges as $\xi\propto\left|\phi-\phi_{c}\right|^{-\nu_{\perp}}.$ 

Remarkably, however, unlike the equilibrium scenario where fluctuations
diverge at the critical point, density fluctuations in these models
are anomalously \emph{small} at criticality\cite{PRL_hyper}, a phenomenon
termed hyperuniformity\cite{torquato_local_2003}. This is seen by
measuring, as a function of $\ell$, the density variance, $\sigma^{2}\left(\ell\right)\equiv\left\langle \delta\rho^{2}\left(\ell\right)\right\rangle $
in a volume $\ell^{d}$. Asymptotically, $\sigma^{2}\left(\ell\right)\sim\ell^{-\lambda}$,
with the exponent $\lambda$ characterizing the magnitude of density
fluctuations. For hyperuniform fluctuations, $\lambda>d$, so that
density fluctuations decay much faster than for a random distribution,
for which $\lambda=d$. The largest possible value of the exponent
$\lambda$ is $d+1$, which occurs for systems like a periodic lattice
\cite{Torq1}. An equivalent measure of hyperuniformity is that the
structure factor $S(k)$ vanishes as $k\rightarrow0,$ typically characterized
by an exponent $\alpha$: $S(k)\sim k^{\alpha}$\cite{zachary_hyperuniformity_2011}.

Although several absorbing state systems have been shown to exhibit
hyperuniformity, the underlying reasons for this unusual behavior
have as yet to be elucidated. In this Letter, we examine a new class
of models with an additional conserved quantity in the dynamics, which
changes behavior in the active phase in an essential way. We then
propose and solve a simple model analytically, and derive a novel
Langevin equation for the coarse-grained dynamics. This leads us to
propose that \emph{large scale hyperuniform fluctuations are determined
by a competition between a noise term which generates fluctuations,
and a deterministic term which reduces them}. We suggest that this
could provide a general mechanism leading to hyperuniformity.

The best-known classes of absorbing state models are the directed
percolation (DP) \cite{DPUni} and the Manna class \cite{MannaUni}.
The difference is that in the Manna class, particle number is conserved,
while in the DP class it is not. Here we study the effect of an additional
conservation law, where the center of mass (COM) of two interacting
particles is conserved by the dynamics, and find that it profoundly
changes the behavior of the system. In colloidal systems at low densities,
COM conservation may emerge naturally at low Reynolds number: if two
spherical particles interact via a repulsive radial force when they
are close, the dynamics will separate them along the line joining
their centers. If more than two particles interact this is no longer
true, but at low densities we expect this to be rare. 

\textcolor{black}{This additional conservation law does not seriously
modify the absorbing phase of the models, but it has a great effect
on the active phase. In the absence of COM conservation, the active
phase of models of the Manna class is characterized by $S(k)\sim k$
for intermediate values of $k$, going over to a constant as $k\rightarrow0$\cite{tjhung2015hyperuniform},
indicating that the system in the active phase is not hyperuniform.
The addition of COM conservation changes this materially, with $S\left(k\right)\sim k^{2}$
as $k\rightarrow0$ in the active phase. This behavior corresponds
to hyperuniform density fluctuations going as $\sigma^{2}\left(\ell\right)\sim\ell^{-(d+1)}$,
which is the fastest decay possible, on par with that of a perfect
lattice. }

After presenting numerical results on two models, we will introduce
and study analytically a simple one-dimensional model, for which we
derive a Langevin equation valid for large densities and long length
scales. While the average profile (averaged over realizations of the
dynamics) obeys a diffusion equation, we find that the additional
conservation law modifies the fashion that noise enters in an essential
way. We find that hyperuniformity emerges through a competition between
diffusion of the average density profile, which reduces density fluctuations,
and noise emerging from the stochastic dynamics, which creates them. 

\begin{figure}
\begin{centering}
\includegraphics[scale=0.4]{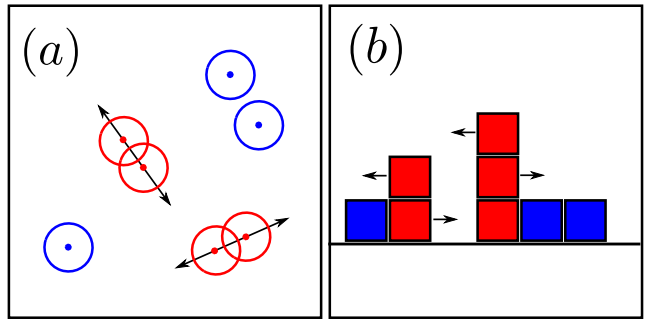}
\par\end{centering}

\protect\caption{(a) An illustration of the random organization model with center of
mass conserving dynamics. (b) The one dimensional toy model. In both
cases blue particles are inactive and red are active. \label{models}}
\end{figure}

We first present numerical results on two variants of models found
in the literature, each modified to include COM conservation%
\footnote{While it is inessential, we take boundary conditions to be periodic
and for the random organization model and Manna model in each step
of the dynamics all are simultaneous so that all active particles
move. %
}:\\
\textbf{I}) Random Organization (2d): Particles of radius $a$ are
placed randomly in an $L\times L$ box. If two particles overlap,
they are considered active and are given a random displacement; otherwise
they are static. We modify the model of \cite{corte} such that the
displacement of a pair of active particles is along the axis connecting
their centers, with an amplitude chosen from a uniform distribution
in the range $\left[0,2a\right]$; see Fig. \ref{models}(a).\\
\textbf{II}) Manna Model (2d): This model\cite{hinrichsen_non-equilibrium_2000}
is defined on a square lattice where each site may hold any number
of particles; sites with more than two particles are considered active.
Two particles on an active site move in opposite directions to adjacent
neighboring site, with vertical or horizontal moves chosen with equal
probability. \\

These models exhibit a phase transition between a low density absorbing
phase and a high density active phase, similar to their forbears without
COM conservation. Moreover, the decay of the density fluctuations
at the critical point is the same as it is for the ordinary Manna-class
models. However, in contrast to the ordinary Manna class, fluctuations
in the active phase of both of these models is hyperuniform. 

Hyperuniformity is readily seen by measuring the structure factor
$S\left(k\right)=\frac{1}{N}\left|\sum_{i=1}^{N}exp\left(-ikr_{i}\right)\right|^{2}$.
Since $S\left(k\rightarrow0\right)=\frac{\left\langle N^{2}\right\rangle -\left\langle N\right\rangle ^{2}}{\left\langle N\right\rangle }$
for an infinite system\cite{torquato_local_2003}, hyperuniformity
at large scales implies that $S\left(k\rightarrow0\right)\rightarrow0$
\footnote{neglecting the delta function at $k=0$%
}\cite{torquato_local_2003,Torq2}, typically%
\footnote{ But not always; see Ref. \cite{BandGap2}%
} as a power of $k$: $S\left(k\right)\propto k^{\gamma}$. When $\gamma<1$
the density fluctuations are given by $\sigma^{2}\left(\ell\right)\propto\ell^{-\gamma-d}$,
while if $\gamma>1$ then the fluctuations are what might be termed
`maximally hyperuniform', with $\sigma^{2}\left(\ell\right)\propto\ell^{-d-1}$
\cite{zachary_hyperuniformity_2011}.

\begin{figure}
\includegraphics[scale=0.6]{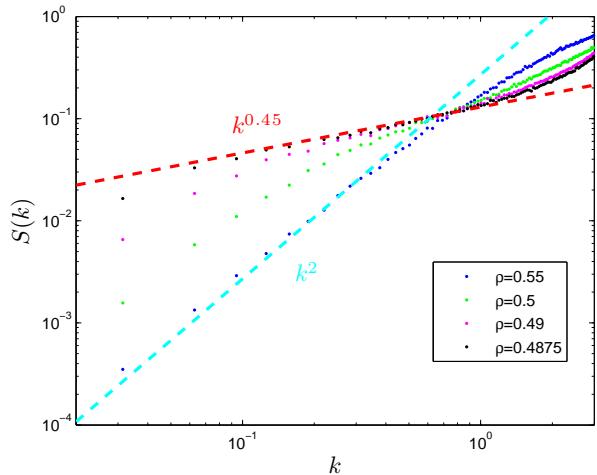}

\protect\caption{2D RandOrg with COM conservation: The structure factor for $\rho>\rho_{c}$
scales as $k^{2}$ as $k\rightarrow0$. Here $L=400$, and the number
of realizations is $50$.\label{fig:RandOrg-1} At the critical point,
$\rho_{c}\approx0.487$, the usual scaling $k^{0.45}$ is seen.}
\end{figure}

\begin{figure}
\begin{centering}
\includegraphics[scale=0.6]{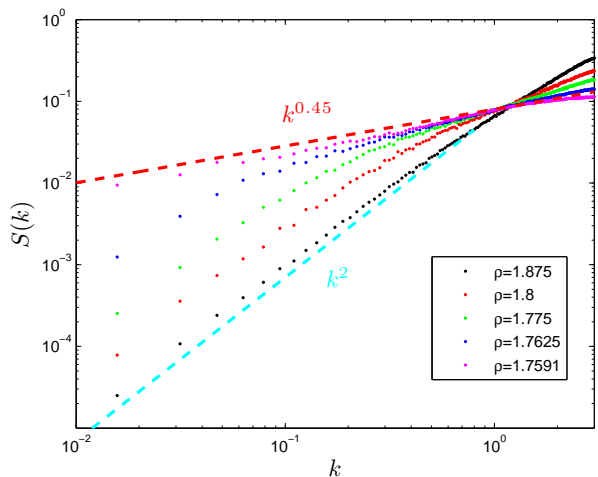}
\par\end{centering}

\protect\caption{2D Manna model with COM conservation for $\rho>\rho_{c}$. Here $L=400$,
and the number of realizations is $50$.\label{fig:Manna}\textcolor{red}{{}
}Here $\rho_{c}\approx1.7591$.}
\end{figure}

As seen in Figures \ref{fig:RandOrg-1} and \ref{fig:Manna}, which
show $S(k)$ above the critical density for the two models described
above, as $k\rightarrow0,$ $S\left(k\right)\propto k^{2}$, implying
that the systems are maximally hyperuniform. As the density approaches
the critical point, a crossover occurs from small values of $k$,
where $S\left(k\right)\propto k^{2}$, to larger $k$, where $S\left(k\right)\propto k^{0.45}$.
We denote the crossover wavevector by $k_{+},$which is most clearly
seen for $\rho=1.7625$ in Figure \ref{fig:Manna}. The large $k$
regime is the same as the \textcolor{black}{critical} \textcolor{black}{behavior}
of absorbing phase transitions without the COM symmetry, as seen in
Reference \cite{PRL_hyper}\textcolor{blue}{. }As the density approaches
its critical value from above, the crossover point $k_{+}$ approaches
$0$, suggesting a diverging correlation length defined as $\xi=2\pi/k_{+}.$

\begin{figure}
\begin{centering}
\includegraphics[scale=0.6]{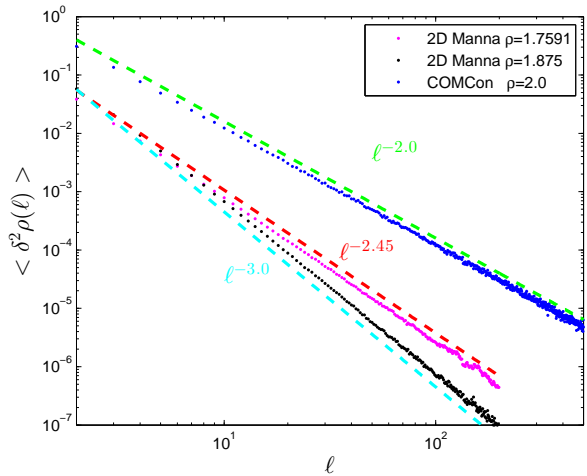}
\par\end{centering}

\protect\caption{Real space fluctuations of the 2D COM Manna model and the COMCon model.
In both cases the density fluctuations at high density is `maximally
hyperuniform', $\sigma^{2}\left(\ell\right)\propto\ell^{-d-1}$. Near
the critical point of the Manna model ($\rho_{c}\approx1.7591$) $\sigma^{2}\left(\ell\right)\propto\ell^{-2.45}$,
similarly to the case where COM symmetry is absent.\label{fig:Manna-real_space}}
\end{figure}

Thus, the central effect of COM conservation is that for all values
of density $\rho>\rho_{c}$ the system becomes hyperuniform, so that
$\frac{\left\langle N^{2}\right\rangle -\left\langle N\right\rangle ^{2}}{\left\langle N\right\rangle }\rightarrow0$
in an \textcolor{black}{infinite} system. This is in contrast to the
case where this symmetry is absent and hyperuniformity occurs only
when the system is tuned to the critical point. Figure \ref{fig:Manna-real_space}
shows this same behavior in real space for the two dimensional Manna
model. Near the critical point, the system shows the usual scaling
$\sigma^{2}\left(\ell\right)\propto\ell^{-2.45}$\cite{PRL_hyper},
while for large densities, $\sigma^{2}\left(\ell\right)\propto\ell^{-3}$
similar to the fluctuations of a crystal whose `atoms' are randomly
displaced from their lattice sites. 

\begin{figure}
\begin{centering}
\includegraphics[scale=0.6]{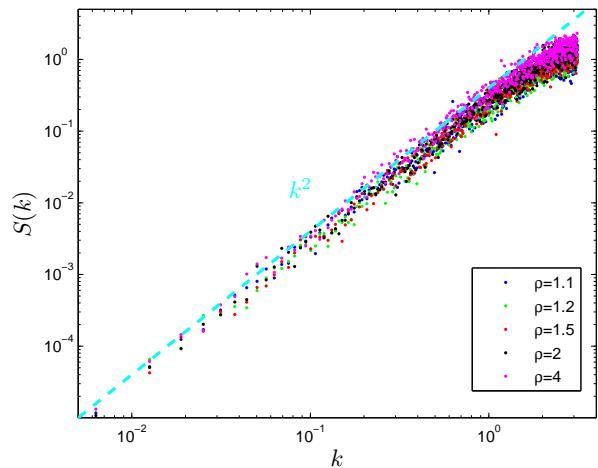}
\par\end{centering}

\protect\caption{The structure factor, $S\left(k\right)\propto k^{2}$, of COMCon for
density $\rho>1$, corresponding to the active phase. The system size
here is $L=1000$ and the number of realizations is $20$ for each
density. \label{fig:Toy}}
\end{figure}

In order to better understand the origins of the observed hyperuniformity,
we now introduce a one-dimensional model, which we call the COMCon
model, which admits analytic analysis in its active phase. In this
model, a site\textcolor{black}{{} $i$ may have any number $n_{i}$
of particles. If $n_{i}>1$, then at a rate $\omega_{0}\cdot\left(n_{i}-1\right)$,
two particles move from site $i$ - one moves to site $i-1$ while
the second moves to $i+1$, thus conserving the center of mass.} The
phase transition for the COMCon model is of a different character
from that of models I and II, as seen in both the structure factor,
which diverges as $S\left(k\right)\sim k^{-0.5}$ as $\rho_{c}\approx1$
is approached from below, and in the fraction of active particles,
which jumps as $\rho$ crosses $\rho_{c}$. However, like these other
two models, $S\left(k\right)\sim k^{2}$ in the active phase of COMCon,
as seen in Figure \ref{fig:Toy}. Here we will only study the behavior
of the model where $n_{i}>>1$, which ensures that the system is everywhere
active. 

Let us begin by studying the coarse grained dynamics of COMCon in
the continuum limit at large densities $\rho>\rho_{c}\approx1$. As
shown in the Supplementary Materials, the Langevin equation for the
continuum density $n\left(x,t\right)$ has a diffusive term and a
noise term emanating from the microscopic random dynamics: 
\begin{equation}
\partial_{t}n=D\,\partial_{xx}n+A\,\partial_{xx}\left(\sqrt{n-1}\,\eta\right)\label{eq:Lang-1}
\end{equation}
Here $\left\langle \eta\left(x,t\right)\eta\left(x',t'\right)\right\rangle =\delta\left(x-x'\right)\delta\left(t-t'\right)$,
$A=\sqrt{\omega_{0}}a^{2}$, and $D=\omega_{0}a^{2}$ are constants
which depend on the microscopic rates, and $a$ is the lattice constant.
The key point here is that due to particle conservation \emph{and}
COM conservation, the noise term comes in as a second spatial derivative.
It is this that leads to the far-reaching consequence of hyperuniformity
in the active phase, as discussed below. 

We note that Equation \ref{eq:Lang-1} only describes the model when
the average density is large, since when $n<1$ we get spurious results
- complex noise and diffusion - which is not present in the discrete
model. Such issues arise because we chose the transition rates to
be $\omega_{0}\left(n-1\right)$, and treat $n$ as a continuous variable.
To avoid this anomalous behavior, we will restrict our analysis to
the high density regime %
\footnote{A possible remedy is to enforce $n>1$ in the transition rates using
a step function $\lambda\left(n-1\right)\Theta\left(n-1\right)$ where
$\Theta\left(x\right)$ is unity for $x>0$ and otherwise zero.%
}, that is, the active phase. 

The second point to note is that if only particles were conserved,
the noise term would have a single space derivative, familiar from
the Model B dynamics\cite{hohenberg1977theory} of ordinary diffusion.
Such a noise term would not lead to hyperuniform fluctuations. 

The easiest way to see that the density fluctuations deriving from
Equation \ref{eq:Lang-1} are hyperuniform is to compute the structure
factor $S\left(k\right)=\frac{1}{N}\left\langle n_{k}n_{-k}\right\rangle $,
where $n_{k}=\int dx\, n\left(x\right)e^{-ikx}$. Writing $n(x)=n_{0}+\delta n$,
where $n_{0}$ is the average density, and expanding to first order
in $\delta n$, the last term in Equation \ref{eq:Lang-1} can be
approximated by $A\,\sqrt{n_{0}-1}\,\partial_{xx}\eta$. Writing the
Langevin in Fourier space, we get

\begin{equation}
\partial_{t}\delta n_{k}=-k^{2}D\delta n_{k}-A\sqrt{n_{0}-1}\, k^{2}\eta_{k}\label{eq:Langevin_k-space}
\end{equation}
whose solution is 

\begin{equation}
\delta n_{k}=Ce^{-Dk^{2}t}+k^{2}A\sqrt{n_{0}-1}e^{-Dk^{2}t}\int_{0}^{t}dt'e^{Dk^{2}t'}\eta_{k}\label{eq:deln}
\end{equation}
where $C$ is set by the initial condition. We may now compute the
$k$ dependence of the structure factor, by using the fact that $S\left(k\right)=\frac{1}{N}\left\langle \delta n_{k}\delta n_{-k}\right\rangle $: 

\begin{eqnarray}
S(k) & = & C^{2}e^{-2Dk^{2}t}+\frac{n_{0}-1}{n_{0}}\frac{A^{2}k^{2}}{2D}\left(1-e^{-2Dk^{2}t}\right)\\
 & \underset{t\rightarrow\infty}{\rightarrow} & \frac{n_{0}-1}{n_{0}}\frac{A^{2}}{2D}k^{2}.
\end{eqnarray}
Thus, the result of COM conservation is that $S(k)\propto k^{2}$. 

This model calculation points up one way that hyperuniformity can
occur. In particular, we note that the two terms of Equation \ref{eq:Langevin_k-space}
have competing roles. The diffusion term reduces fluctuations so that
a given mode with wave vector $k$ decays with a rate $-k^{2}D$ for
small $k$. On the other hand the second term on the right hand side
generates fluctuations which scale as $k^{2}\,\eta_{k}$, which are
greatly suppressed at small $k$ values.

In conclusion we have shown that COM leads to hyperuniformity emerging
as a competition between a diffusion term suppressing a fluctuations
and a noise term which generates fluctuations. The interplay between
the scaling with $k$ of these two terms determines the steady state
fluctuations. It is interesting to speculate that other hyperuniform
systems are governed by a similar Langevin equation, perhaps with
different scalings. For example, for a Langevin equation (expressed
in Fourier space) of the form 
\begin{equation}
\partial_{t}n_{k}=-D\left|k\right|^{\alpha}n_{k}-Ak^{\beta}\eta_{k}\label{eq:Gener-1}
\end{equation}
 the structure factor scales as $S\left(k\right)=\left|k\right|^{2\beta-\alpha}$.
Typically, if the noise has no special properties, $\beta=1$, and
in order for hyperuniformity to occur we would need $\alpha<2$. The
exponent $\alpha$ determines the time it takes the system to reach
steady state, which corresponds to the decay time of the longest mode
$k=\frac{2\pi}{L}$. From Equation \ref{eq:Gener-1} it is found that
that $\tau\propto L^{\alpha}$ so that if $\alpha<2$ a steady state
is reached much faster than diffusion where $\tau\propto L^{2}$.
Faster than diffusion scaling is found at the critical point of the
Manna universality class \cite{Lubek} for dimensions smaller than
four which have been shown to be hyperuniform \cite{PRL_hyper}. 

We thank Paul Chaikin and Yariv Kafri for interesting and helpful
discussions. DL thanks the U.S. - Israel Binational Science Foundation
(Grant No. 2014713), the Israel Science Foundation (Grant No. 1254/12),
and the Russell Berrie Nanotechnology Institute, Technion, and the
Initiative for the Theoretical Sciences of the Graduate Center, CUNY.

\subsection{Appendix A: Direct derivation of Langevin equation \label{sec:Direct-derivation}}

In this section we derive the coarse grained Langevin equation for
the COMCon lattice model in one dimension. Towards this goal we first
show that center-of-mass (COM) conservation law constrains the current
to be a divergence of a field. Particle conservation implies that
the change in the density can be written as a divergence of the current:

\begin{equation}
\partial_{t}n=-\nabla\cdot J.\label{eq:cont_eq}
\end{equation}
Thus, by the divergence theorem, the change in the particle number
within any volume must come from the boundary. 

As a result of COM conservation each dimension can be associated with
global conserved quantities $R_{\alpha}=\int d^{d}r\, n\left(r\right)r_{\alpha}$,
so that $\partial_{t}R_{\alpha}=0$, where the integral is over all
space. If the integral is taken over a finite portion of the system,
any change in $R_{\alpha}$ is due to particles entering or exiting
along the surface enclosing this region. This can be expressed as
$\partial_{t}R_{\alpha}=-\int dS\cdot J_{R,\alpha}$ where $J_{R,\alpha}$
is the current associated with the COM conservation. Using Eq. \ref{eq:cont_eq}
the change in $R_{\alpha}$ can be computed, 
\begin{eqnarray}
\partial_{t}R_{\alpha} & = & \int d^{d}r\, r_{\alpha}\partial_{t}n\left(r\right)\\
 & = & -\int d^{d}r\, r_{\alpha}\nabla\cdot J\\
 & = & -\int dS\cdot\left(Jr_{\alpha}\right)+\int d^{d}rJ_{\alpha}
\end{eqnarray}
where in the last line we integrated by parts. COM conservation requires
the last term to also scale as the surface, implying that $J_{\alpha}$
can be written as the divergence of a vector, $J_{\alpha}=-\nabla\cdot\sigma_{\alpha}$.
Together with Equation \ref{eq:cont_eq}, this gives

\begin{equation}
\partial_{t}n=\nabla\cdot\nabla\cdot\sigma_{\alpha}.\label{eq:Sigma}
\end{equation}
In one dimension, $\sigma$ may be written $\sigma=f\left(n\right)+g\left(n\right)\eta$,
being composed of a deterministic term $f\left(n\right)$ and a term
$g\left(n\right)\eta$ which accounts for the stochastic motion, where
$\eta$ is assumed to be zero averaged white noise $\left\langle \eta\left(x,t\right)\eta\left(x',t'\right)\right\rangle =\delta\left(x-x'\right)\delta\left(t-t'\right)$. 

Since $\left\langle \eta\right\rangle =0$, the term $f\left(n\right)$
can be found by computing $\left\langle \Delta n\right\rangle =\int_{0}^{\Delta t}dt\left\langle \partial_{t}n\right\rangle =\partial_{xx}f\left(n\right)$.
The noise term $g\left(n\right)$ can be found by looking at $\left\langle \left(\Delta n\right)^{2}\right\rangle $,
since in an infinitesimal duration, the noise term $\int_{0}^{\Delta t}dt\, g\left(n\right)\eta\sim\sqrt{\Delta t}$,
which dominates over the deterministic term which scales as $\Delta t$.
For the COMCon model, two particles exit a site at a rate $\omega_{0}\left(n_{i}-1\right)$,
and move in opposite directions to adjacent sites. The average change
in the occupancy $\left\langle \Delta n_{i}\right\rangle $ is composed
of three terms, where the first is due to the aforementioned transition
and the other two account for transitions in neighboring sites. Hence, 

\begin{eqnarray}
\left\langle \Delta n_{i}\right\rangle  & = & \left[-2\left(n_{i}-1\right)+\left(n_{i+1}-1\right)+\left(n_{i-1}-1\right)\right]\omega_{0}\Delta t\nonumber \\
 & \approx & \omega_{0}a^{2}\Delta t\frac{\partial^{2}n}{\partial^{2}x}\label{eq:mean_DN}
\end{eqnarray}
where we took the continuum limit by assuming that sites are separated
by a lattice constant $a$. Therefore, $f(n)=\omega_{0}a^{2}n.$ Eq.
\ref{eq:mean_DN} implies that the average density profile evolves
via usual diffusion, with no alterations due to COM conservation.

To find $g\left(n\right)$ we compute $\left\langle \left(\Delta n\right)^{2}\right\rangle $
since as noted, the fluctuations dominate, allowing us to neglect
any deterministic terms. This is calculated from the term $\partial_{xx}g\left(n\right)\eta$.
For simplicity, we discretize space, taking the lattice constant to
be $a$: 
\begin{equation}
\Delta n_{i}=\int_{0}^{\Delta t}dt\frac{1}{a^{2}}\left[g\left(n_{i+1}\right)\eta_{i+1}+g\left(n_{i-1}\right)\eta_{i-1}-2g\left(n_{i}\right)\eta_{i}\right].
\end{equation}
Assuming that $\left\langle \eta_{i}\eta_{j}\right\rangle =\delta_{ij}$,
the second moment is then given by

\begin{equation}
\left\langle \left(\Delta n_{i}\right)^{2}\right\rangle =\frac{1}{a^{4}}\left[g^{2}\left(n_{i+1}\right)+g^{2}\left(n_{i-1}\right)+4g^{2}\left(n_{i}\right)\right]\Delta t.
\end{equation}
 This is compared to the exact result for the COMCon model to order
$O\left(\Delta t\right)$, 

\begin{equation}
\left\langle \left(\Delta n_{i}\right)^{2}\right\rangle =\left[\left(n_{i-1}-1\right)+\left(n_{i+1}-1\right)+4\left(n_{i}-1\right)\right]\omega_{0}\Delta t,
\end{equation}
which implies that $g\left(n\right)=\sqrt{\omega_{0}a^{4}}\sqrt{n-1}$.
The main result of this section is the Langevin equation, given by, 

\begin{equation}
\partial_{t}n=\omega_{0}a^{2}\partial_{xx}n+\sqrt{\omega_{0}a^{4}}\partial_{xx}\sqrt{n-1}\eta\label{eq:Langevin}
\end{equation}

\subsection{Appendix B: Field theory derivation of Langevin equation }

In this section we present an alternative derivation of the coarse
grained Langevin equation for the COMCon model. Towards this goal,
we employ the framework of Ref. \cite{Biroli} to  compute its field
theory and then take the (spatial) continuum limit. Let $n_{i}\left(t\right)$
denote the number of particles at site $i$ and at time $t$ which
is discritized into intervals of $dt$. We first compute the probability
measure $P\left(\left\{ n_{i}\left(t\right)\right\} \right)$ of a
given trajectory $\left\{ n_{i}\left(t\right)\right\} $, which we
express as a functional integral over an auxiliary field $\left\{ p_{i}\left(t\right)\right\} $;
this will allow us to identify the action $\mathfrak{s}\left(\left\{ n_{i}\right\} ,\left\{ p_{i}\right\} \right)$:
\begin{equation}
P\left(\left\{ n_{i}\left(t\right)\right\} \right)=\frac{1}{Z}\int\mathcal{D}p\: e^{-\mathfrak{s}\left(\left\{ n_{i}\right\} ,\left\{ p_{i}\right\} \right)}\label{eq:P1}
\end{equation}
where $\mathcal{D}p\equiv\Pi dp_{j}(t)$ and the product over all
lattice sites and over all times in the interval of interest. The
action can be identified by expressing the average of an arbitrary
functional $\left\langle O\left[\left\{ n_{i}\left(t\right)\right\} \right]\right\rangle $
over some period of time as a path integral over all possible trajectories,
where the dynamics of the allowed trajectories is enforced using delta
functions. We write these dynamics as $\Delta n_{i}(t)\equiv n_{i}\left(t+dt\right)-n_{i}\left(t\right)=J_{i}$,
where the `currents' $J_{i}$ reflect the occupancy changes permitted
by the rules, and then average over all values of $J_{i}$ :

\begin{eqnarray}
\left\langle O\right\rangle  & = & \frac{1}{Z}\left\langle \int\mathcal{D}n\: O\left[\left\{ n_{i}\left(t\right)\right\} \right]\Pi_{i}\delta\left(\Delta n_{i}(t)-J_{i}\right)\right\rangle _{J},
\end{eqnarray}
where $\Pi_{i}$ represents a product over all independent currents
and times.

For the COMCon, the COM conservation condition gives $J_{i-1}=J_{i+1}=1$
and $J_{i}=-2$. Using the Martin-Siggia-Rose (MSR) procedure \cite{MSR1,MSR2,MSR3},
the delta functions are written as integrals over plane waves:

\begin{equation}
\left\langle O\right\rangle =\frac{1}{Z}\int\mathcal{D}n\mathcal{D}p\: O\left[\left\{ n_{i}\left(t\right)\right\} \right]\left\langle \Pi_{t}e^{-\sum_{i}\, p_{i}\left(\Delta n_{i}(t)-J_{i}\right)}\right\rangle _{J}
\end{equation}
where $\Pi_{t}$ is a product over discretized time. Hence the probability
density of a given trajectory is

\begin{eqnarray}
P\left(\left\{ n_{i}\left(t\right)\right\} \right) & = & \frac{1}{Z}\int\mathcal{D}p\left\langle \Pi_{t}e^{-\sum_{i}p_{i}\left(\Delta n_{i}(t)-J_{i}\right)}\right\rangle _{J}\label{eq:P2}
\end{eqnarray}
where we identify the action $\mathfrak{s}$ by comparing Equations
\ref{eq:P1} and \ref{eq:P2}. In this expression, the only model-dependent
quantity is the generating function, $\left\langle e^{\sum_{i}p_{i}J_{i}}\right\rangle $.
For the COMCon model, 
\begin{eqnarray}
\left\langle e^{\sum_{i}p_{i}J_{i}}\right\rangle  & = & \Big(1-\omega_{0}\sum_{i}\left(n_{i}-1\right)dt\Big)\\
 & +\omega_{0} & \sum_{i}\left(n_{i}-1\right)dt\: e^{\Delta^{2}p_{i}}\\
 & \approx & e^{\omega_{0}\sum_{i}\left(n_{i}-1\right)dt\left(e^{\Delta^{2}p_{i}}-1\right)}
\end{eqnarray}
where $\Delta^{2}p_{i}\equiv p_{i+1}-2p_{i}+p_{i-1}$. The first term
is the probability that no transition occurs during time $dt$, and
the second term sums over the probabilities of a single transition.
The probability of two transitions occurring is of order $\left(dt\right)^{2}$
and is therefore neglected. Reverting to continuum time by taking
$\sum dt\rightarrow\int dt$ which leads to the final action $\mathfrak{s}$
given by: 

\begin{equation}
\mathfrak{s}=\sum_{i}\int dtp_{i}\partial_{t}n_{i}-\omega_{0}\left(n_{i}-1\right)\left(e^{\Delta^{2}p_{i}}-1\right).
\end{equation}
To obtain the continuum limit we assume that adjacent sites are separated
by a small distance $a$ and expand the second term, so that $p_{i\pm1}\approx p_{i}\pm a\partial_{x}p_{i}+\frac{1}{2}a^{2}\partial_{xx}p_{i}\pm\frac{a^{3}}{3!}\partial_{x}^{\left(3\right)}p_{i}+\frac{a^{4}}{4!}\partial_{x}^{\left(4\right)}p_{i}$:

\begin{eqnarray}
e^{\Delta^{2}p_{i}}-1 & \approx & e^{a^{2}\partial_{xx}p+2a^{4}/4!\partial_{x}^{4}p_{i}}-1\\
 & \approx & a^{2}\partial_{xx}p+\frac{2a^{4}}{4!}\partial_{x}^{4}p+\frac{1}{2}a^{4}\left(\partial_{xx}p\right)^{2}\label{eq:expansion}
\end{eqnarray}
The second term in Eq. \ref{eq:expansion} subdominant since it has
more spatial derivatives than the first term and henceforth neglected.
The continuum action is then given by:

\begin{eqnarray}
\mathfrak{s} & = & -\int dt\int dx[p\partial_{t}n-\omega_{0}a^{2}p\partial_{xx}n-\label{eq:Action_1}\\
 &  & \,\,\,\,\,\,\,\,\,\,\,\,\,\,\,\,\,\,\omega_{0}\left(n-1\right)\frac{1}{2}a^{4}\left(\partial_{xx}p\right)^{2}]\nonumber 
\end{eqnarray}
where the second term is obtained through integration by parts. The
Langevin equation associated with this action is identified by a writing
a trial Langevin equation with different possible terms and then performing
the MSR procedure to obtain the resulting action. Motivated by Section
\ref{sec:Direct-derivation} we assume

\[
\partial_{t}n=D\partial_{xx}n+\partial_{xx}\xi
\]
where $\xi(x,t)$ is a Gaussian noise term and its dependence on $n_{i}$
is to be determined. To obtain the action we express the average of
any quantity $O\left[\left\{ n_{i}\left(t\right)\right\} \right]$as
a path integral over all possible trajectories, where the dynamics
are enforced by delta functions. Subsequently, the delta function
written in terms of the auxiliary field $p$, 

\begin{eqnarray*}
P\left\langle O\right\rangle  & = & \frac{1}{Z}\left\langle \int\mathcal{D}n\: O\left[\left\{ n_{i}\left(t\right)\right\} \right]\delta\left(\partial_{t}n-D\partial_{xx}n+\partial_{xx}\xi\right)\right\rangle _{\xi}\\
 & = & \frac{1}{Z}\left\langle \int\mathcal{D}n\mathcal{D}p\, O\left[\left\{ n_{i}\left(t\right)\right\} \right]e^{-\int dt\int dx\, p\,\left(\partial_{t}n-D\partial_{xx}n-\partial_{xx}\xi\right)}\right\rangle _{\xi}.
\end{eqnarray*}
The first two terms are identical those in Eq. \ref{eq:Action_1},
implying that the diffusion constant $D=\omega_{0}a^{2}$. The angler
brackets denote an average over the Gaussian noise $\xi$, with distribution
$p\left(\left\{ \xi_{i}\right\} \right)\propto e^{-\int dx\int dt\frac{\xi^{2}}{2g^{2}\left(n\right)}}$
where we $g\left(n\right)$ is to be found. Only the last term depends
on $\xi$ and its average can be computed by discretizing space and
time and then performing the Gaussian integral over $\xi$ in a period
of $dt$. This yields, 
\begin{eqnarray}
 &  & \left\langle e^{-\int dxdtAp\partial_{xx}\xi}\right\rangle _{\xi}=\\
 & = & \int\Pi_{i}d\xi_{i}\frac{1}{\sqrt{2\pi g^{2}\left(n_{i}\right)}}e^{-\sum_{i}dt\frac{A}{a^{2}}p_{i}\Delta^{2}\xi_{i}-\sum_{i}\frac{\xi_{i}^{2}}{2g^{2}\left(n_{i}\right)}dt}\\
 & = & exp\left[\sum_{i}\frac{1}{2}\left(\frac{A}{a^{2}}\Delta^{2}p_{i}\right)^{2}g^{2}\left(n_{i}\right)dt\right]\\
 & \approx & exp\left[\int dxdt\frac{1}{2}\left(\frac{1}{a^{2}}\partial_{xx}p\right)^{2}g^{2}\left(n\right)\right]
\end{eqnarray}
where the continuum limit has been taken in the last line. Here, $\Delta^{2}\xi_{i}=\xi_{i+1}-2\xi_{i}+\xi_{i-1}$
is the second (spatial) difference of $\xi$, and likewise $\Delta^{2}p_{i}$.
Comparing to Eq. \ref{eq:Action_1} we see that $g^{2}\left(n\right)=\omega_{0}a^{4}\left(n-1\right)$
in agreement with Eq. \ref{eq:Langevin}.

\bibliographystyle{unsrt}
\bibliography{CM}

\end{document}